\documentclass[fleqn,10pt]{wlscirep}
\usepackage[utf8]{inputenc}
\usepackage[T1]{fontenc}
\usepackage{amsmath,amssymb,amsthm}
\usepackage{graphicx}
\usepackage{multirow}
\usepackage{subcaption}
\usepackage{hyperref}
\usepackage{xcolor}
\usepackage{wasysym}
\newcommand{\RNum}[1]{\uppercase\expandafter{\romannumeral #1\relax}}

\title{VIRIS: Simulating indoor airborne transmission combining architectural design and people movement}

\author[1,2]{Yidan Xue}
\author[3]{Wassim Jabi}
\author[1]{Thomas E. Woolley}
\author[1,*]{Katerina Kaouri}
\affil[1]{Cardiff University, School of Mathematics, Cardiff, CF24 4AG, UK}
\affil[2]{The University of Manchester, School of Health Sciences, Manchester, M13 9PL, UK}
\affil[3]{Cardiff University, Welsh School of Architecture, Cardiff, CF10 3NB, UK}

\affil[*]{KaouriK@cardiff.ac.uk}


\begin{abstract}
A Viral Infection Risk Indoor Simulator (VIRIS) has been developed to quickly assess and compare mitigations for airborne disease spread. This agent-based simulator combines people movement in an indoor space, viral transmission modelling and detailed architectural design, and it is powered by topologicpy, an open-source Python library. VIRIS generates very fast predictions of the viral concentration and the spatiotemporal infection risk for individuals as they move through a given space. The simulator is validated with data from a courtroom superspreader event. A sensitivity study for unknown parameter values is also performed. We compare several non-pharmaceutical interventions (NPIs) issued in UK government guidance, for two indoor settings: a care home and a supermarket. Additionally, we have developed the user-friendly VIRIS web app that allows quick exploration of diverse scenarios of interest and visualisation, allowing policymakers, architects and space managers to easily design or assess infection risk in an indoor space.
\end{abstract}
\begin{document}

\flushbottom
\maketitle

\thispagestyle{empty}

\section{Introduction}

In the COVID-19 pandemic, more than 700 million were infected and more than 7 million died \cite{covid_data}. One reason for the rapid and widespread transmission is that, unlike many other infectious diseases, it can be transmitted via aerosols \cite{Asadi2020}. The aerosols can remain suspended in the air for hours and can travel long distances \cite{Wang2021}, thus, increasing infection risk.

During the pandemic, policymakers and space managers had to make fast decisions to mitigate transmission, often without available scientific inputs. Mathematical modellers, including some of the authors here, collaborated closely with governments and developed models of transmission to inform policy, for various indoor scenarios including educational settings \cite{Lau2022,Moore2021}, restaurants \cite{LiYuguo2021,Lau2022,Balkan2024}, supermarkets \cite{Xu2020,Ying2021,Cui2021}, a choral practice \cite{Miller2021}, a courtroom \cite{Vernez2021,Lau2022} and a train carriage \cite{deKreij2022}. These models quantify the infection risk in indoor spaces for different non-pharmaceutical interventions (NPIs).

However, people movement was considered in very few models \cite{Xu2020,Ying2021,Zhen2022,Ciunkiewicz2022,Luca2022,Balkan2024}, limiting the accuracy of simulations, especially in settings where people are actively moving around (e.g.~supermarkets). In addition, most models either neglect the physics of the spread of infectious aerosols \cite{Xu2020,Ying2021,Vernez2021,Miller2021,Ciunkiewicz2022,Luca2022}, or use computational fluid dynamics simulations \cite{LiYuguo2021,Cui2021,Zhen2022}, making them too computationally intensive for fast policymaking. Here, building on research Kaouri and Woolley used to inform policy during the pandemic \cite{Lau2022,Moore2021}, we present VIRIS, a new simulator and a web app that incorporates both people movement and aerosol physics in geometrically complex, architecturally defined spaces, providing solutions with great speed. Following Lau et al.~\cite{Lau2022}, we use a reaction-diffusion equation to simulate the spatiotemporal distribution of infectious aerosols and determine the spatiotemporal infection risk extending the Wells-Riley ansatz \cite{Riley1978}. People movement is modelled using an agent-based approach based on a navigation graph, embedded into topologicpy \cite{topologic,topologicpy} for the purpose of this work. Topologicpy is an advanced spatial modelling and analysis software (Python) library for architecture, engineering, and construction; further details on topologicpy and the navigation graph can be found in a companion paper \cite{Wassim2024}.

In Section \ref{sec:methodology}, we describe the modelling framework underpinning VIRIS and discuss parameter values. In Section \ref{sec:results}, we demonstrate how VIRIS can be used to model a series of diverse NPIs, in a care home and in a supermarket. In addition, in Section \ref{sec:web_app}, we introduce a user-friendly web app underpinned by VIRIS. The results are discussed in Section \ref{sec:discussion}.

\section{Mathematical model and assumptions}
\label{sec:methodology}
We present VIRIS, an agent-based modelling framework and simulator that combines airborne viral transmission, people movement and detailed architectural design. Below, we present the model, assumptions, parameter values, and its numerical implementation.

\subsection{Assumptions}

An infectious individual is assumed to be a moving point source of infectious aerosols. We assume that the aerosols are emitted with zero velocity, a good assumption, since we do not consider coughing or sneezing \cite{Bourouiba2014}. In reality, particle size varies, following a probability density function \cite{Johnson2011,Bagheri2023}; for simplicity, and similarly to other works \cite{LiYuguo2021,Lau2022,Balkan2024}, we assume
that the aerosols can be modelled using a single representative size, with a constant gravitational settling rate \cite{DeOliveira2021}, $\gamma$. Furthermore, we are justified to neglect large droplets, since evaporation happens very quickly (seconds to minutes) \cite{Wells1934,Xie2007}.

Experimental studies \cite{Dabisch2021a,Aganovic2021} show that the virus deactivation rate depends on a variety of factors including relative humidity, temperature and sunlight. From data \cite{Dabisch2021a,Aganovic2021}, we assume a constant virus deactivation rate, $\beta$, as reported by van Doramalen et al.~\cite{VanDoremalen2020}; thus is an accurate assumption for most indoor settings (temperature around 20$^{\circ}$C and relative humidity below 60\%).

The infectious aerosols are not well-mixed in an indoor space \cite{Bhagat2020,Bhagat2024}; they assume a spatiotemporal distribution. Here, we focus on single storey buildings, but our methodology can be extended to multistorey buildings. The indoor space, which includes furniture, is bounded by walls or doors. Furthermore, without loss of generality, we assume perfect reflection of aerosols off walls \cite{Lau2022}. Furniture that is higher than a person's height are also modelled as walls; otherwise we assume that furniture has no effect on viral transmission, since the airborne particles can flow over them. We assume that the infectious aerosols are well-mixed along the room height; this allows to reduce the three-dimensional (3D) model to a two-dimensional (2D) model of viral transmission. 

Poor ventilation has been observed in many diverse indoor settings (for example, schools \cite{Guo2008}, care homes \cite{Srikanth2024}, supermarkets \cite{LiChunying2021}, courtrooms \cite{Vernez2021}, and restaurants \cite{LiYuguo2021}), usually when windows are closed and fresh air is mainly provided by a mechanical ventilation system ineffectively. For these poorly ventilated spaces, we can neglect the advective transport of aerosols due to air flows \cite{Bhagat2020,Bhagat2024}. However, in well ventilated spaces, air flows play a more significant role in viral transmission \cite{Cui2021,Zhen2022}. Therefore, applying our model to well ventilated settings requires further validation with real-world data. We represent ventilation as a first-order removal of aerosols using a constant air exchange rate, $\lambda$. To quantify ventilation quality, we use air changes per hour (ACH), which is widely used by building engineers and space managers (ACH is converted to $\lambda$ via $1\ \mathrm{ACH} \approx 2.78\times10^{-4}$ air changes per second).

Following the Wells-Riley ansatz \cite{Wells1955,Riley1978}, we, thus, model the first-order removal of infectious aerosols with rate $\kappa$, which is a sum of the virus deactivation rate, gravitational settling rate, and ventilation rate, respectively, that is
\begin{equation}
\kappa=\beta+\gamma+\lambda.
\label{eq:decay}
\end{equation}
All parameter values are given in Table \ref{table:parameter_values}. 

We approximate the diffusion coefficient of aerosols using the turbulent eddy diffusivity, an empirical function that considers local turbulent mixing \cite{Foat2020}:
\begin{equation}
D = 0.8\lambda{V}^{2/3},
\label{eq:diffusion}
\end{equation}
where $V$ is the room volume. We assume a constant diffusion coefficient in each scenario we consider below, determined using the volume of the largest room. Movement of individuals might enhance the advective and diffusive transport of aerosols, as shown in the experiments of Mingotti et al.~\cite{Mingotti2020}, which would then cause a faster spread of the aerosols. Also, the effects of walking speed, crowd density and human-to-source distance on the diffusion coefficient have been investigated in the experiments of Lim et al.~\cite{Lim2024}; some of these effects could be incorporated in future work.

Each individual has several attributes: (\RNum{1}) whether they are infectious or susceptible, (\RNum{2}) whether they are a superspreader, (\RNum{3}) if they wear a mask, (\RNum{4}) their walking speed, $v$, (\RNum{5}) their movement schedule. Each individual follows a schedule consisting of a list of events (e.g.~``watching TV'' or ``attending a lecture''), and each event is characterised by a location $(x,y)$ in a 2D plane, a start time, an end time, and an activity (resting, talking, talking loudly, exercising, and intensive exercising). Furthermore, each activity is characterised by a breathing rate and an aerosol emission rate; see Table \ref{table:parameter_values}.

We assume that an individual is either stationary or moving with a constant speed from an event to the next event, and that each individual always moves ahead of the end time of their current event to arrive exactly at the start time of the next event. We assume that each individual moves along the shortest path between the previous and next event locations (assuming they know the space well), and cannot move through domain boundaries and obstacles, such as walls and furniture. The trajectory $(x_i(t),y_i(t))$ of the $i$th individual is computed in Section \ref{sec:numerics}. Moreover, the infection risk of a susceptible individual depends on the total number of inhaled infectious aerosols (dose) \cite{Riley1978,Dabisch2021b}; we assume that each individual needs the same dose to get infected.

\subsection{Modelling viral aerosols using a reaction-diffusion equation}
Based on the modelling assumptions above, a reaction-diffusion equation governs the concentration of infectious aerosols, $C(x,y,t)$, in the indoor space, which is our domain $\Omega$:
\begin{equation}
\frac{\partial{C}}{\partial{t}}=D\nabla^2C+S-(\kappa+R_I){C},
\label{eq:governing_equation}
\end{equation}
where $S$ is the aerosol source term due to the aerosol emissions from the infectious individuals, $R_I$ is the removal rate of infectious aerosols due to the breathing of all individuals, and $\nabla=(\partial/\partial{x},\partial/\partial{y})$, since we reduce the model to 2D. Perfect reflection of aerosols off walls corresponds to the Neumann boundary conditions $\partial{C}/\partial{n}=0$ on $\partial\Omega$, where $n$ is the outward normal direction.

For $m$ infectious individuals, the aerosol source is
\begin{equation}
S(x,y,t) = \sum_{i=1}^m \frac{(1-\eta_i)R_i}{h}\delta(x-x_i(t))\delta(y-y_i(t)),
\end{equation}
where $\eta$ is the mask efficiency, $R$ is the infectious aerosol emission rate, $h$ is the room height, $\delta$ is the Dirac delta function representing a point source, and the subscript $i$ denotes the parameter value for the $i$th agent. All $n$ individuals (both infectious and susceptible) due to breathing correspond to a sink for the infectious aerosols
\begin{equation}
R_I(x,y,t) = \sum_{i=1}^n (1-\eta_i)\rho_i\delta(x-x_i(t))\delta(y-y_i(t)),
\end{equation}
where $\rho$ is the breathing rate. The infectious aerosol emission rate, $R$, is calculated by $R={\mu}R_t$, where $\mu$ is the viral load, which depends on the virus type and whether the infectious agent is a superspreader, and $R_t$ is the aerosol emission rate, which depends on the activity. The viral load is estimated via
\begin{equation}
\mu=\frac{\pi}{6}d_p^3c_v,
\label{eq:viral_load}
\end{equation}
where $d_p$ is the average aerosol diameter, and $c_v$ is the viral load concentration (copies/mL) \cite{Anand2020}.

\subsection{Spatiotemporal infection risk}

Following the Wells-Riley model \cite{Wells1955,Riley1978}, we assume the infection risk of a susceptible individual is an exponential function of the dose, $d$, the number of inhaled infectious aerosols as follows,
\begin{equation}
P(d)=1-e^{-Id},
\label{eq:infection_risk}
\end{equation}
where $I$ is infectibility, which depends on the median infectious dose $d_m$ via $I=-\ln(0.5)/d_m$. The median infectious dose is the number of viral copies required to infect 50\% of the individuals; the median dose depends on the disease \cite{Qiu2023}. 

The total number of inhaled aerosols can be calculated by
\begin{equation}
d(t)=\int_0^t(1-\eta){\rho}C(x(\tau),y(\tau),\tau)d\tau.
\label{eq:total_aerosols}
\end{equation}
Our model is a discrete S-I-R (Susceptible-Infectious-Removed) model \cite{Allen1994}, where R refers to the recovery from the disease or removal of individuals from the space. Since the simulation runs for less than one day, recovery is not considered; removal takes place in some settings, e.g.~in a supermarket, where infected individuals enter and then leave the space.

\subsection{Parameter values}
\label{sec:parameter_values}

Table \ref{table:parameter_values} summarises the parameters used in the simulations, for the SARS-CoV-2 virus. The value of $c_v$ for the SARS-CoV-2 virus ranges from $10^2$ to $10^{11}$ copies/mL \cite{Anand2020}; the range of $10^8$--$10^9$ copies/mL is most commonly used in modelling studies \cite{Srinivasan2021,Bagheri2021,Buonanno2020,Vernez2021}. Here, we take $c_v=10^9\ \mathrm{copies/mL}$ ($\mu=6.5\%$) for a ``normal'' infectious individual and $c_v=5\times10^9\ \mathrm{copies/mL}$ ($\mu=32.5\%$) for a superspreader \cite{Lau2022}; this was validated with data from a courtroom superspreader outbreak \cite{Vernez2021}. 

For the SARS-CoV-2 virus, the median infectious dose has been found to be 52 and 256 for seroconversion and fever, respectively, using an animal model \cite{Dabisch2021b}. Following Lau et al.~\cite{Lau2022}, we use $d_m=100$. The values of $c_v$ and $d_m$ will be further justified using uncertainty quantification in Section \ref{sec:courtroom}. We consider three types of masks: cotton, surgical and N95, with efficiency 50\%, 60\% and 95\%, respectively \cite{Lindsley2021}. We set the efficiency of N95 mask to 95\% instead of 99\% reported in Lindsley et al.~\cite{Lindsley2021}, since the improper, or repeated use of a mask may reduce its efficiency. The breathing rate and the aerosol emission rate of an individual is based on their activity, which has been provided in Table \ref{table:parameter_values}. When an individual is moving from one event to another, they are assumed to walk with speed 1.5 m/s \cite{Bosina2017}. The parameter values for walking and exercising are assumed to be the same.

\begin{table}
\begin{center}
  \begin{tabular}{cccc} \hline
   Parameter & Symbol & Value & References \\ \hline \rule{0pt}{2.5ex}
    Virus deactivation rate & $\beta$ & $1.7{\times}10^{-4}\ \mathrm{s^{-1}}$ & \cite{VanDoremalen2020,Aganovic2021}\\
    Gravitational settling rate & $\gamma$ & $1.1{\times}10^{-4}\ \mathrm{s^{-1}}$ & \cite{DeOliveira2021}\\
    Air exchange rate & $\mathrm{ACH}(\lambda)$ & Very poor: $\mathrm{ACH}=0.12\ \mathrm{h^{-1}}$ ($\lambda=3.3\times10^{-5}\ \mathrm{s^{-1}}$) & \cite{Guo2008,Lau2022}\\   
    & & Poor: $\mathrm{ACH}=0.72\ \mathrm{h^{-1}}$ ($\lambda=2.0\times10^{-4}\ \mathrm{s^{-1}}$) & \cite{Guo2008,Lau2022}\\
    & & Good: $\mathrm{ACH}=3\ \mathrm{h^{-1}}$ ($\lambda=8.3\times10^{-4}\ \mathrm{s^{-1}}$) & \cite{Guo2008,Lau2022}\\
    Diffusion coefficient & $D$ & $4.6\times10^{-4}$--$4.7\times10^{-2}\ \mathrm{m^{2}/s}$ & \cite{Foat2020}\\
    Viral copies per aerosol & $\mu$ & $6.5\%$--$32.5\%$ & \cite{Anand2020,Buonanno2020,Vernez2021}\\
    Average aerosol diameter & $d_p$ & $5\ \mathrm{\mu{m}}$ & \cite{LiYuguo2021,Lau2022}\\
    Viral load concentration & $c_v$ & $10^9$--$5\times10^9\ \mathrm{copies/mL}$ & \cite{Anand2020,Lau2022}\\
    Median infectious dose & $d_m$ & 100 particles & \cite{Dabisch2021b,Lau2022}\\ \hline \rule{0pt}{2.5ex}
    Breathing rate & $\rho$ & resting: $1.8{\times}10^{-4}\ \mathrm{m^{3}/s}$ & \cite{Orton2022,Mutsch2022} \\
    & & talking: $2.2{\times}10^{-4}\ \mathrm{m^{3}/s}$ & \cite{Orton2022} \\
    & & talking loudly: $2.5{\times}10^{-4}\ \mathrm{m^{3}/s}$ & \cite{Orton2022} \\
    & & exercising: $1.1{\times}10^{-3}\ \mathrm{m^{3}/s}$ & \cite{Berry1996,Orton2022} \\
    & & intensive exercising: $2.2{\times}10^{-3}\ \mathrm{m^{3}/s}$ & \cite{Mutsch2022} \\ \hline \rule{0pt}{2.5ex}
    Aerosol emission rate & $R_t$ & resting: 8 $\mathrm{particles/s}$ & \cite{Asadi2019,Lau2022}\\
    & & talking: 40 $\mathrm{particles/s}$ & \cite{Asadi2019,Lau2022}\\
    & & talking loudly: 80 $\mathrm{particles/s}$ & \cite{Asadi2019,Lau2022}\\
    & & exercising: 145 $\mathrm{particles/s}$ & \cite{Orton2022}\\
    & & intensive exercising: 625 $\mathrm{particles/s}$ & \cite{Orton2022}\\ \hline \rule{0pt}{2.5ex}
    Mask efficiency & $\eta$ & cotton: 50\% & \cite{Lindsley2021}\\
    & & surgical: 60\% & \cite{Lindsley2021}\\
    & & N95: 95\% & \cite{Lindsley2021}\\ \hline \rule{0pt}{2.5ex}
    Walking speed & $v$ & 1.5m/s & \cite{Bosina2017}\\
\hline
  \end{tabular}
\end{center}
\caption{Parameter values used in the simulations, for the SARS-CoV-2 virus. Values for $\beta$, $\mu$, $c_v$ and $d_m$ are different for other viruses.}
\label{table:parameter_values}
\end{table}

\subsection{Numerical implementation}
\label{sec:numerics}

The domain $\Omega$ and the architectural design are constructed using topologicpy \cite{topologic,topologicpy}, which has been extended to incorporate people movement for the purposes of this project. Topologicpy is a Python-based spatial modelling and topological analysis library that supports the conceptual design of architectures; details are given in a companion paper \cite{Wassim2024}. The first step is to create cuboids to represent rooms and corridors. The surfaces of the cuboids represent walls, floors or ceilings. Furniture is also modelled by cuboids. To simulate the movement of individuals, we generate a navigation graph, $G$, for the indoor space \cite{Wassim2024}. The vertices of this graph are the locations of all possible events, and the edges connecting the vertices represent all possible pathways an individual can move on. For each two neighbouring events, we compute on $G$ the shortest path between the event locations \cite{Wassim2024}. Each individual moves with constant walking speed, and arrive exactly at the start time of the next event. 

The reaction-diffusion equation (\ref{eq:governing_equation}) is solved on $\Omega$ using a finite element method (FEM) \cite{Langtangen2017}. We generate a mesh for $\Omega$ using the default Frontal-Delaunay algorithm in the gmsh library \cite{gmsh}. Then we discretise the time derivative using a backward Euler scheme, discretise the space using piecewise linear elements and compute the FEM problem at each time step using the Python-based open-source FEM library scikit-fem \cite{skfem}. Note that using scikit-fem enables smooth integration with topologicpy in a lightweight fashion, since scikit-fem is smaller than 1MB in size and only uses NumPy \cite{numpy} and SciPy \cite{scipy} libraries, which are used in topologicpy \cite{topologicpy}. VIRIS runs very fast; for example, for tens of individuals with a one-day schedule, it only takes seconds to compute the concentration and the infection risk on a standard laptop. All Python scripts can be found in a GitHub repository at \url{https://github.com/KaterinaKaouri/VIRIS}.

\section{Results}
\label{sec:results}

In this section, we apply the simulator to three different indoor settings---courtroom, care home, and supermarket---to demonstrate the applicability of the model and its ability to model diverse NPIs. At the end, we present the VIRIS app, a user-friendly web app underpinned by the VIRIS simulator.

\subsection{Courtroom}
\label{sec:courtroom}

We consider a courtroom case study from Vaud, Switzerland, in October 2020. The locations, schedules and outcomes of all individuals are described in detail \cite{Vernez2021}. Three to four individuals were infected during a 3-hour court hearing \cite{Vernez2021}. This case study has also been modelled in Lau et al.~\cite{Lau2022}, and it was shown that the infectious individual has to be a superspreader.

\begin{figure}
  \begin{subfigure}[b]{0.4\columnwidth}
    \centering
    \includegraphics[width=\linewidth]{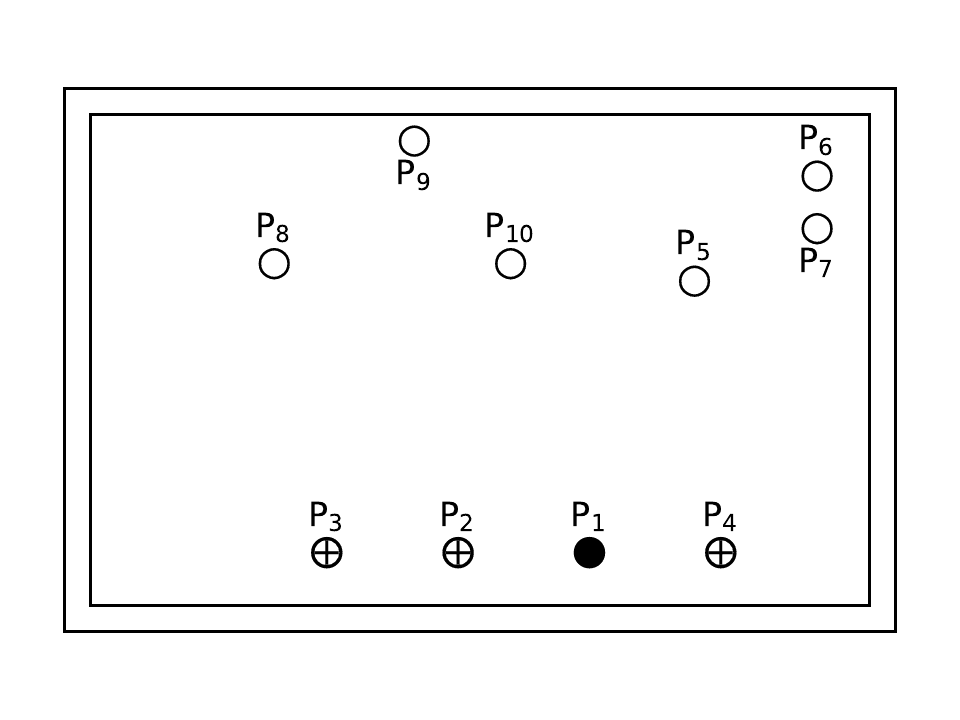}
    \caption{Courtroom schematic ($\CIRCLE$ for infectious, $\bigoplus$ for infected, and $\bigcirc$ for not infected)}
    \subfloat[The activities of each individual]{%
    \footnotesize
    \begin{tabular}{cccc}
      \hline
      Individual & Activity (hearing) & Activity (breaks)\\ \hline
   $\mathrm{P_{1}}$ & talking & talking \\
   $\mathrm{P_{2}}$ & talking loudly & talking \\
   $\mathrm{P_{3}}$ & resting & talking \\
   $\mathrm{P_{4}}$ & talking & talking \\
   $\mathrm{P_{5}}$ & talking loudly & N/A \\
   $\mathrm{P_{6}}$ & resting & N/A \\
   $\mathrm{P_{7}}$ & resting & N/A \\
   $\mathrm{P_{8}}$ & talking loudly & N/A \\
   $\mathrm{P_{9}}$ & resting & N/A \\
   $\mathrm{P_{10}}$ & resting & N/A \\
    \end{tabular}
  }
  \end{subfigure}
  \hfill 
  \begin{subfigure}[b]{0.59\columnwidth}
    \includegraphics[width=\linewidth]{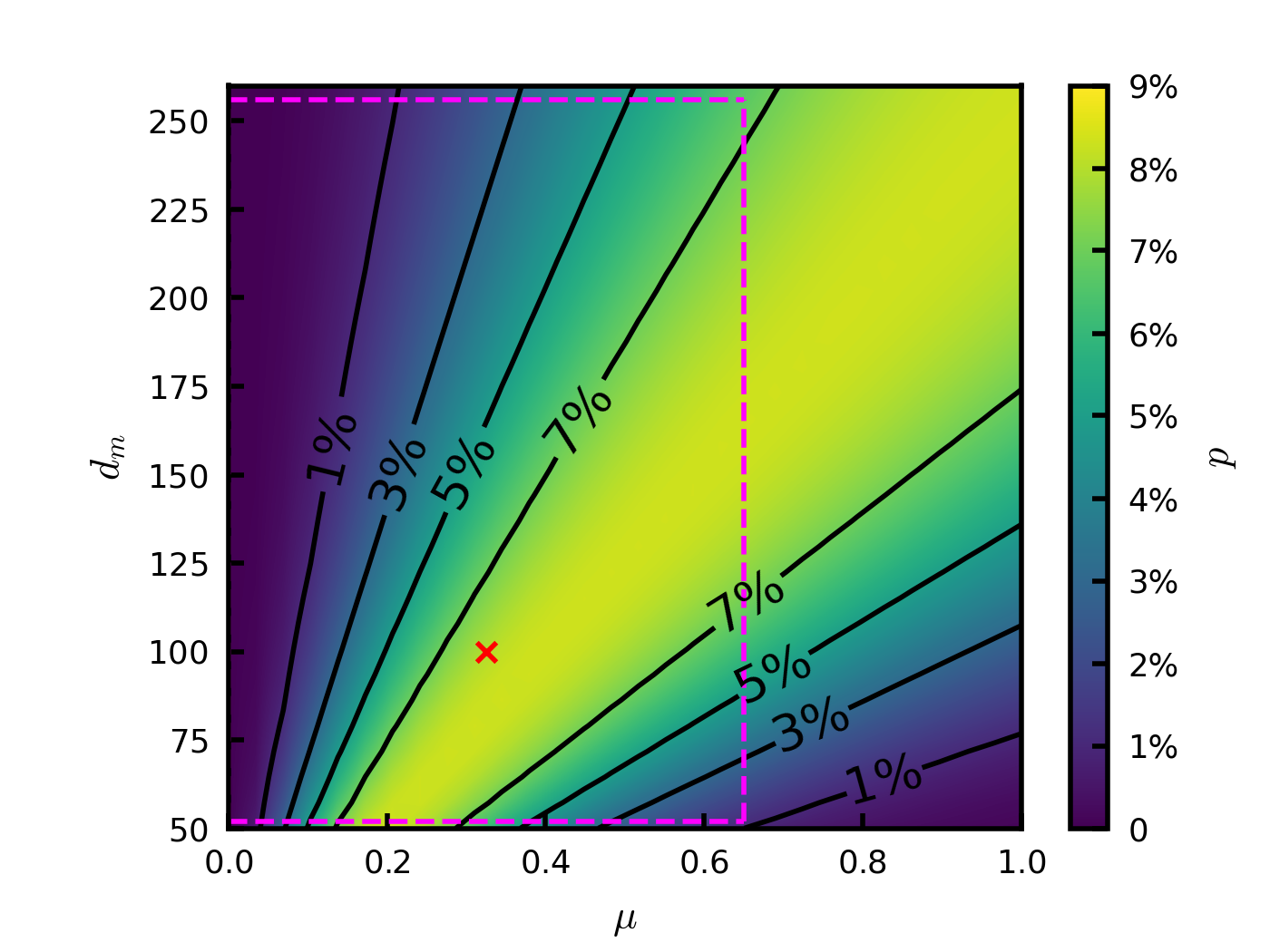}
    \caption{Probability of reproducing the data (uncertainty quantification)}
  \end{subfigure}
  \caption{Courtroom scenario: (a) the location of each individual, and their infection status ($\CIRCLE$ for infectious, $\bigoplus$ for infected, and $\bigcirc$ for not infected) \cite{Vernez2021}. (b) The activities of each individual during the hearing and the breaks \cite{Vernez2021}. (c) Probability of the model reproducing (a), calculated by equation (\ref{eq:probability}), as $\mu$ varies from $0$ to $100\%$ and $d_m$ varies from $50$ to $260$. The red cross indicates the parameter values used in our work. The purple dashed box encloses the possible values \cite{Anand2020,Buonanno2020,Vernez2021,Dabisch2021b}.}
  \label{fig:p_value}
\end{figure}

The courtroom had dimensions of $8.9\ \mathrm{m}\ (l)\times5.6\ \mathrm{m}\ (w)\times3\ \mathrm{m}\ (h)$, so the room volume is 150 $\mathrm{m^3}$. The room was very poorly ventilated, with $\mathrm{ACH}=0.23\ \mathrm{h^{-1}}$ ($\lambda=6.4\times10^{-5}\ \mathrm{s^{-1}}$). Using equation (\ref{eq:diffusion}), the diffusion coefficient is determined to be $D=1.4\times10^{-3}\ \mathrm{m^2/s}$. The hearing involved 10 individuals ($\mathrm{P_1}$--$\mathrm{P_{10}}$), with locations as shown in Figure \ref{fig:p_value}a.

The court hearing lasted from 2:00pm to 5:00pm. There were breaks from 2:00pm to 2:05pm, from 2:23pm to 2:30pm, from 2:55pm to 3:10pm, and from 3:44pm to 3:50pm, during which only the jury members $\mathrm{P_1}$--$\mathrm{P_4}$ remained in the room \cite{Vernez2021}. The witness $\mathrm{P_{10}}$ was only present from 3:10pm to 3:44pm. As shown in Figure \ref{fig:p_value}a, $\mathrm{P_1}$ was the only infectious individual, and is assumed to be a superspreader with $\mu=32.5\%$. $\mathrm{P_2}$, $\mathrm{P_3}$, $\mathrm{P_4}$ and $\mathrm{P_5}$ were found infected after the hearing. However, $\mathrm{P_5}$ had contact with another infectious individual before this hearing, so we assume that $\mathrm{P_5}$ was not infected during the hearing. The activities of each individual during the hearing are listed in Figure \ref{fig:p_value}b. All other parameters are as in Table \ref{table:parameter_values}.

We perform a parameter sweep for $\mu$ and $d_m$ to examine which pair of parameter values results in the highest probability,
\begin{equation}
p = \prod_{i=2}^{4}P_i\prod_{i=5}^{10}(1-P_i),
\label{eq:probability}
\end{equation}
of reproducing exactly the infection data reported in Vernez et al.~\cite{Vernez2021} ($\mathrm{P_2}$--$\mathrm{P_4}$ infected and $\mathrm{P_5}$--$\mathrm{P_{10}}$ not infected). Figure \ref{fig:p_value} visualises $p$ for $\mu$ ranging from 0 to 100\% and $d_m$ ranging from 50 to 260. The purple dashed box encloses possible values of these parameters \cite{Anand2020,Buonanno2020,Vernez2021,Dabisch2021b}. Note that we allowed $\mu$ to be an order of magnitude greater than the upper limit used in recent modelling studies \cite{Buonanno2020,Vernez2021}, but still within the range found in experimental work \cite{Anand2020}. The red cross indicates the parameter values used in this paper, which falls within the region where we cannot reject the null hypothesis, that our model exactly reproduces the reality since $p>0.05$.

\begin{figure}
  \centering
   \begin{subfigure}{\textwidth}
     \centering
     \includegraphics[width=.7\textwidth]{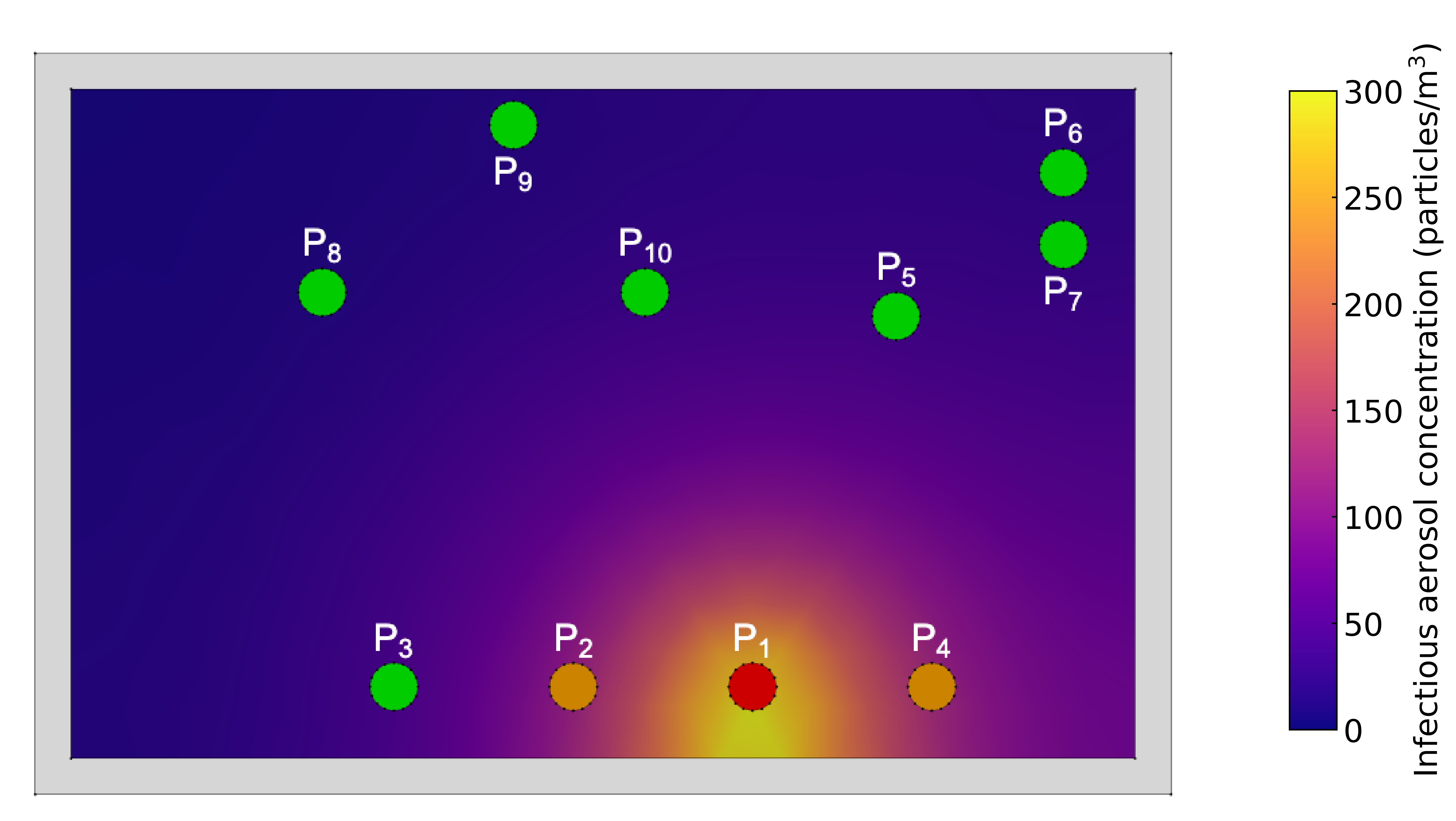}
     \caption{Infectious aerosol concentration}
 \end{subfigure}
  \begin{subfigure}{\textwidth}
     \centering
     \includegraphics[width=.7\textwidth]{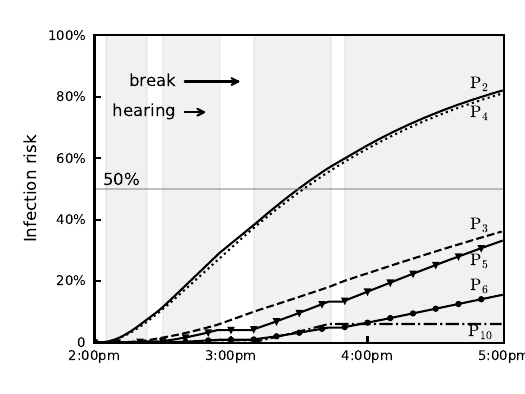}
     \caption{Infection risk for each susceptible individual}
 \end{subfigure}
  \caption{Courtroom scenario: (a) the location of each individual, their infection status (red colour for infectious, orange colour for infected, and green colour for not infected), and the concentration of infectious aerosols at the end of the hearing (5:00pm). The threshold of infection risk was set at 50\% for an individual to get infected. (b) The infection risk of susceptible individuals, $\mathrm{P_2}$, $\mathrm{P_3}$, $\mathrm{P_4}$, $\mathrm{P_5}$, $\mathrm{P_6}$, $\mathrm{P_{10}}$. Individuals $\mathrm{P_7}$, $\mathrm{P_8}$ and $\mathrm{P_9}$ are omitted from this plot, since their infection risk curve is low (below 50\%), and approximately equal to that of $\mathrm{P_6}$.}
  \label{fig:courtroom}
\end{figure}

Figure \ref{fig:courtroom}a shows the concentration of infectious aerosols at the end of the hearing (5:00pm). Figure \ref{fig:courtroom}b shows the infection risk of each susceptible individual over time, where orange colour represents the individuals who are infected after the hearing, while green colour represents who are not. We set the threshold of infection risk at 50\% to determine whether a susceptible individual is infected. Note that a different threshold will result in a different number of infected people (e.g.~$\mathrm{P_3}$ will also be found to be infected, if we set the threshold at 35\%).

As shown in Figure \ref{fig:p_value}a, $\mathrm{P_2}$ and $\mathrm{P_4}$ were closest to the infectious individual $\mathrm{P_1}$ (at 1.5 m) during the entire 3-hour hearing, where a high infectious aerosol concentration and infection risk is predicted by the model. $\mathrm{P_3}$ was also present in the entire hearing but was slightly further away from $\mathrm{P_1}$ (at 3 m); $\mathrm{P_3}$, thus, has the third highest infection risk. $\mathrm{P_5}$ and $\mathrm{P_{10}}$ were at a similar distance from $\mathrm{P_1}$; $\mathrm{P_5}$ has a much higher infection risk than $\mathrm{P_{10}}$, as $\mathrm{P_{10}}$ was only present in the courtroom for only 34 minutes. This result confirms that both proximity and duration of contact have significant impact on the infection risk.

\subsection{Care home}

Next, we consider a typical section of a UK care home \cite{Khaliq2024}, where six individuals ($\mathrm{P_1}$--$\mathrm{P_6}$) move around from 8:00am to 11:00pm. One of the individuals ($\mathrm{P_1}$) is infectious. This scenario demonstrates the versatility of our model to consider different agent schedules and NPIs. Many of the NPIs we study below were suggested by the UK government during the pandemic \cite{care_home_guidance}.

\begin{figure}
  \centering
   \begin{subfigure}[b]{.7\textwidth}
     \centering
     \includegraphics[width=\textwidth]{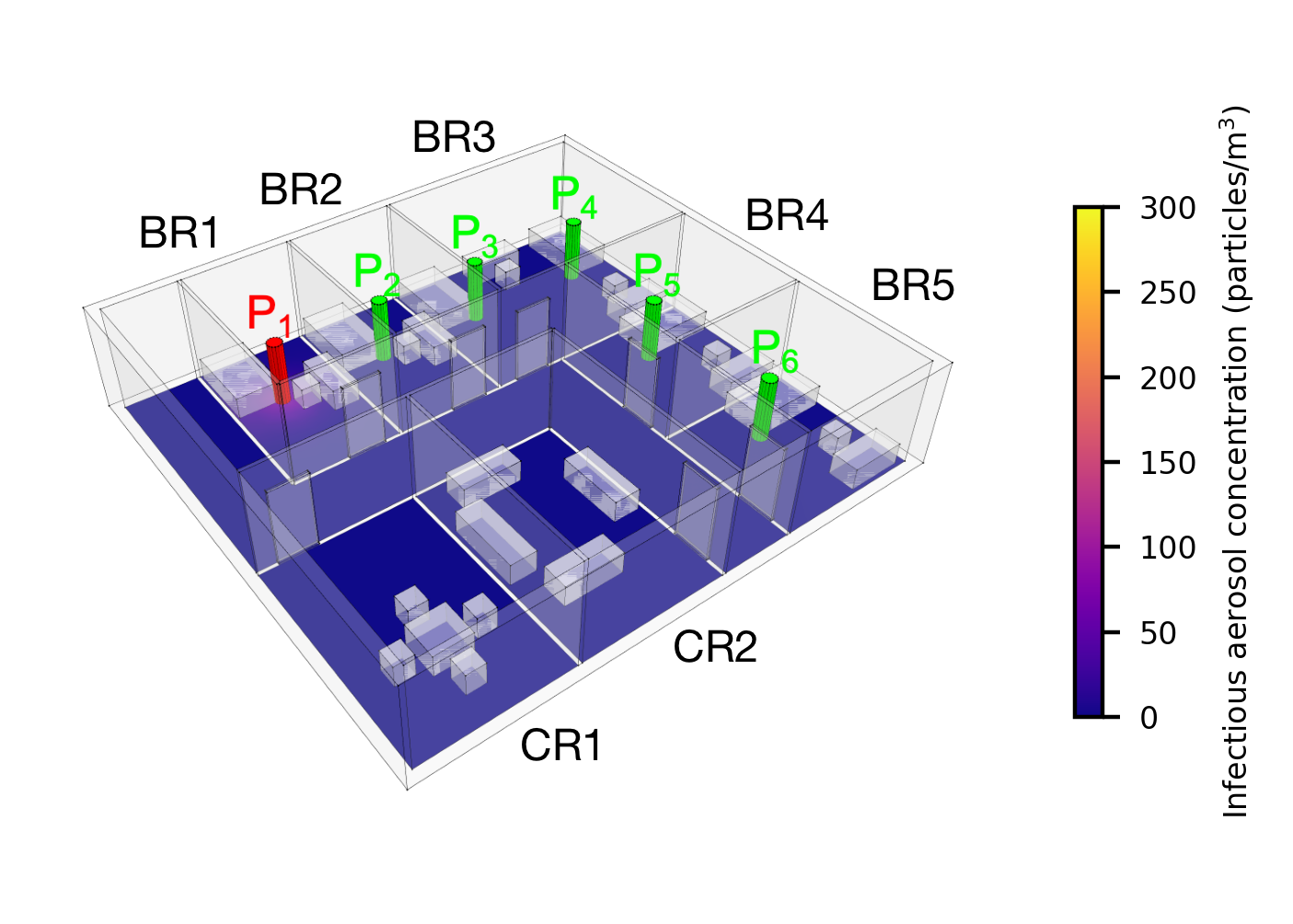}
     \caption{8:30am}
 \end{subfigure}
 \begin{subfigure}[b]{.7\textwidth}
  \centering
  \includegraphics[width=\textwidth]{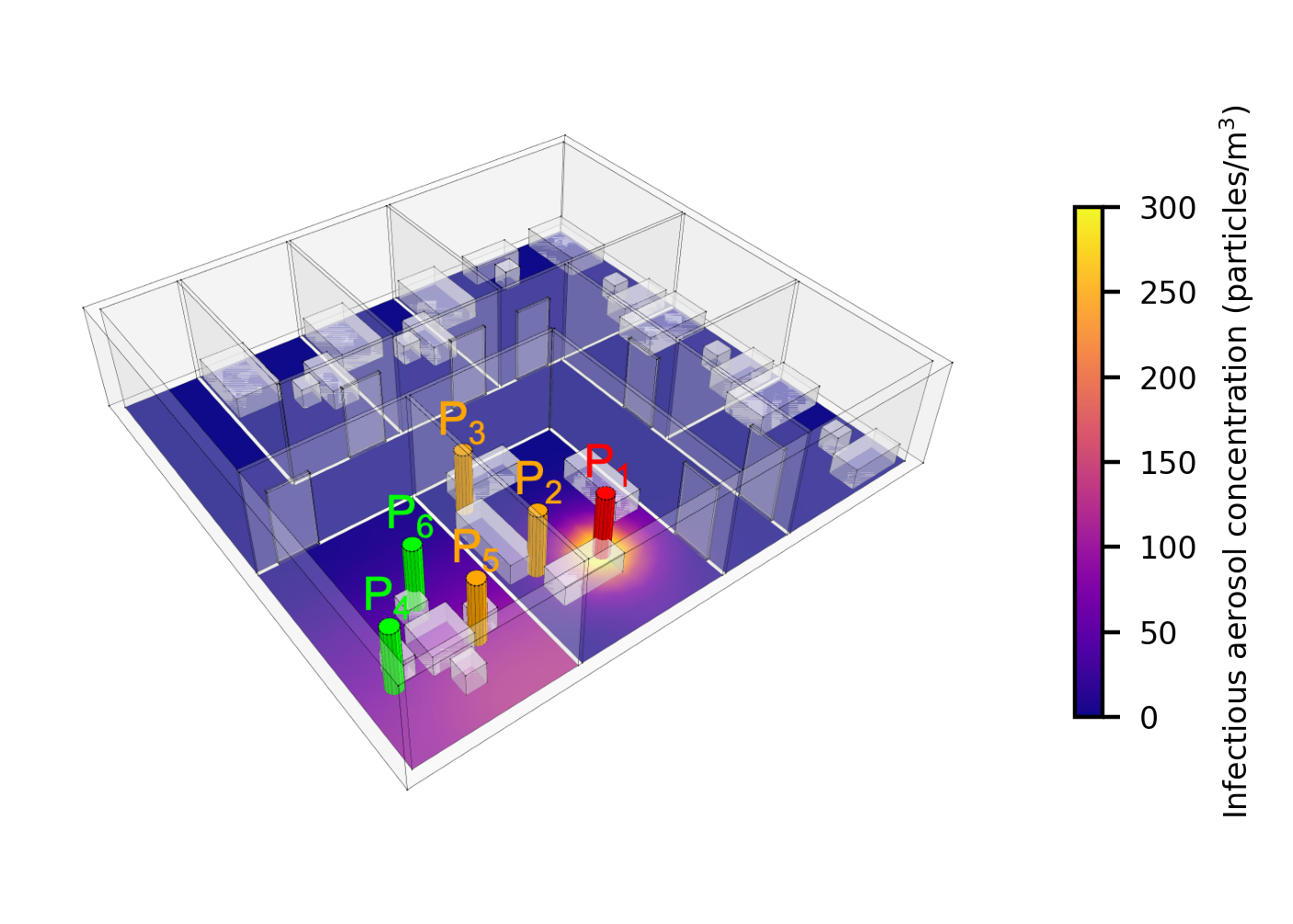}
  \caption{5:30pm}
\end{subfigure}
  \caption{Care home scenario. Each individual is represented by a cylinder and gives a colour according to their infection status. The concentration of infectious aerosols is represented using a colour scale. Two snapshots of the simulation are shown at (a) 8:30am (all individuals are resting in their bedrooms) and (b) 5:30pm. At 5:30pm, three more individuals are likely to be infected.}
  \label{fig:care_home}
\end{figure}

The care home has dimensions of $14\ \mathrm{m}\ (l)\times13\ \mathrm{m}\ (w)\times3\ \mathrm{m}\ (h)$, and includes five bedrooms (BRs) and two common rooms (CRs), as shown in Figure \ref{fig:care_home}. We assume that each individual is in one of the two common rooms from 9:00am to 12:00pm ($\mathrm{P_1}$, $\mathrm{P_3}$, $\mathrm{P_5}$ in CR1; $\mathrm{P_2}$, $\mathrm{P_4}$, $\mathrm{P_6}$ in CR2), from 2:00pm to 5:00pm ($\mathrm{P_1}$, $\mathrm{P_2}$, $\mathrm{P_3}$ in CR1; $\mathrm{P_4}$, $\mathrm{P_5}$, $\mathrm{P_6}$ in CR2), and from 5:00pm to 8:00pm ($\mathrm{P_4}$, $\mathrm{P_5}$, $\mathrm{P_6}$ in CR1; $\mathrm{P_1}$, $\mathrm{P_2}$, $\mathrm{P_3}$ in CR2). At any other time, each individual is in their bedroom (see Figure \ref{fig:care_home}a). Each individual is assumed to be resting when in the bedroom and talking in a common room.

We assume $\mathrm{P_1}$ is a superspreader ($\mu=32.5\%$) and that the ventilation is very poor ($\lambda=3.3{\times}10^{-5}\ \mathrm{s^{-1}}$, $\mathrm{ACH}=0.12\ \mathrm{h^{-1}}$). We set $D=4.6\times10^{-4}\ \mathrm{m^2/s}$ using the common room volume $V=72\ \mathrm{m^3}$ in equation (\ref{eq:diffusion}). We model the movement of individuals based on the prescribed schedule and simulate the concentration of infectious aerosols in the common room and the infection risk of each individual. Figure \ref{fig:care_home}b presents the infectious aerosol concentration and the six individuals' locations at 5:30pm; we see that three individuals ($\mathrm{P_2}$, $\mathrm{P_3}$, $\mathrm{P_5}$) have already gotten infected by $\mathrm{P_1}$.

We compare six NPIs: (\RNum{1}) improving ventilation from very poor ($\mathrm{ACH}=0.12\ \mathrm{h^{-1}}$) to good ($\mathrm{ACH}=3\ \mathrm{h^{-1}}$), (\RNum{2}) everyone wears a surgical mask ($\eta=60\%$), (\RNum{3}) no common room time, (\RNum{4}) reducing the time in the common room by 50\%, (\RNum{5}) introducing a two-hour break with no access to the common room from 5:00pm to 7:00pm, and delaying the third common room gathering to 7:00pm to 10:00pm (\RNum{6}) individuals only interact within 2 small social groups of three. Interventions \RNum{1}, \RNum{3} and \RNum{4} correspond to the UK government's guidance on infection prevention and control measures for social care providers and managers, as updated in July, 2024 \cite{care_home_guidance}. The NPIs we choose showcase how easy is to customise our model to assess diverse interventions, and provide useful comparisons. For instance, we design intervention \RNum{5} because we observe that $\mathrm{P_1}$ leaves behind them a high concentration of infectious aerosols; individuals entering the space after $\mathrm{P_1}$ inhale the aerosols (see the purple regions in CR1 of Figure \ref{fig:care_home}b) and, thus, experience higher infection risk.

\begin{figure}
  \centering
  \includegraphics[width=.7\textwidth]{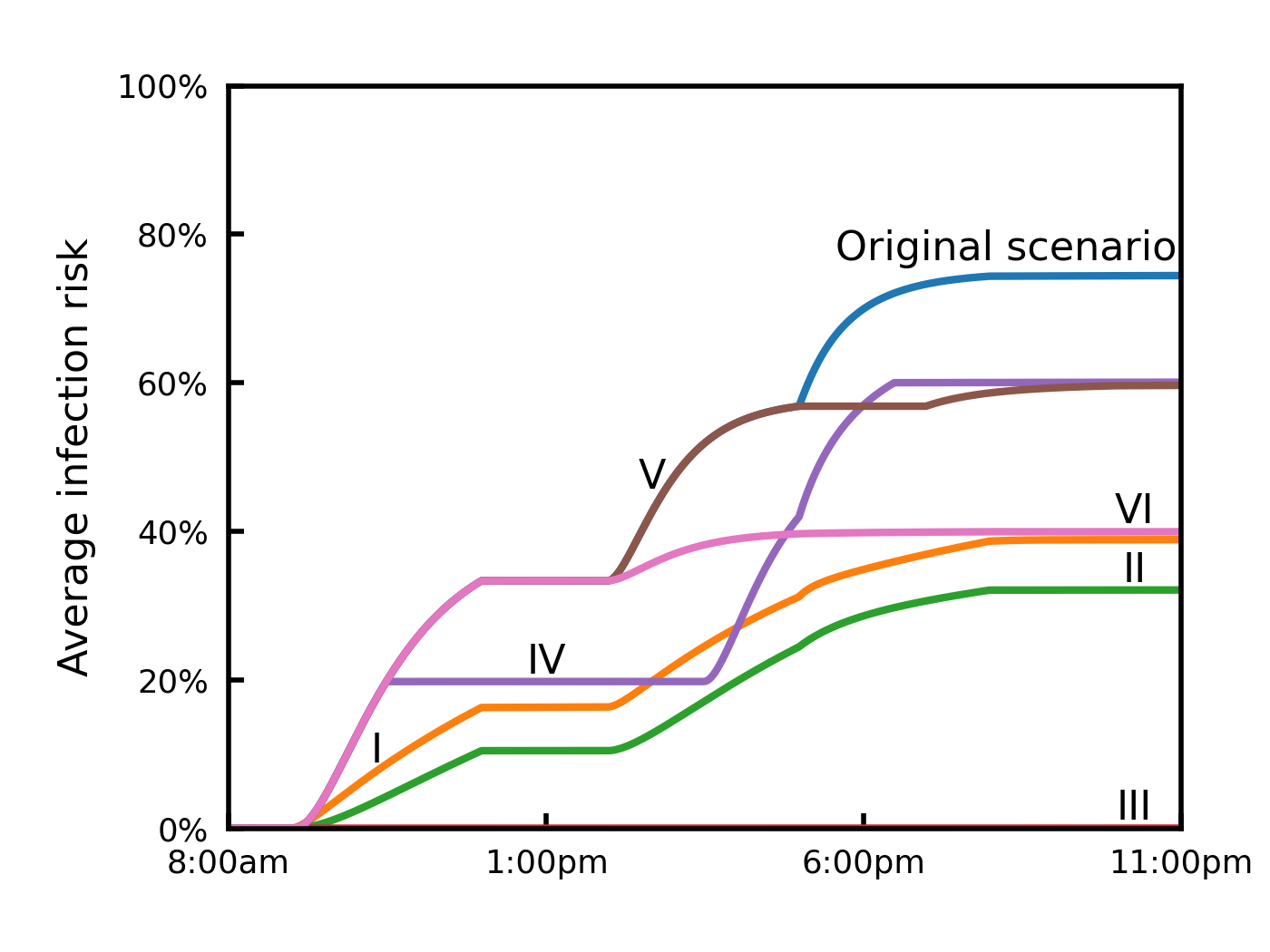}
  \caption{The average infection risk of the five susceptible individuals in the original care home scenario, and when implementing six different NPIs: (\RNum{1}) improving ventilation from very poor ($\mathrm{ACH}=0.12\ \mathrm{h^{-1}}$) to good ($\mathrm{ACH}=3\ \mathrm{h^{-1}}$), (\RNum{2}) everyone wears a surgical mask ($\eta=60\%$), (\RNum{3}) no common room time, (\RNum{4}) reducing common room time by 50\%, (\RNum{5}) introducing a two-hour break with no access to the common room from 5:00pm to 7:00pm, and moving the third common room event to 7:00pm to 10:00pm (\RNum{6}) individuals only interact within two small social groups of three.}
  \label{fig:schedule_comparison_care_home}
\end{figure}

In Figure \ref{fig:schedule_comparison_care_home}, we present the average infection risk of the five susceptible individuals $\mathrm{P_2}$--$\mathrm{P_6}$ in the original scenario, and when implementing interventions \RNum{1}--\RNum{6}. The average infection risk is calculated as the sum of infection risk of all susceptible individuals, divided by the number of susceptible individuals. The most effective intervention is, as expected, \RNum{3} (no common room time), where the infection risk is zero, since there is no contact between the infectious individual and others. The second most effective intervention is \RNum{2} (everyone wears a surgical mask), which reduces the average infection risk by more than 40\%. The third most effective intervention is \RNum{1} (improving ventilation), followed by intervention \RNum{6} (individuals only interact within two small social groups of three), which also results in a reduction of more than 30\% in the average infection risk. The other two interventions (\RNum{4}, \RNum{5}) show similar effectiveness and reduce the average infection risk by approximately 15\%.

\subsection{Supermarket}

\begin{figure}
  \centering
  \begin{subfigure}[b]{.99\textwidth}
     \centering
     \includegraphics[width=\textwidth]{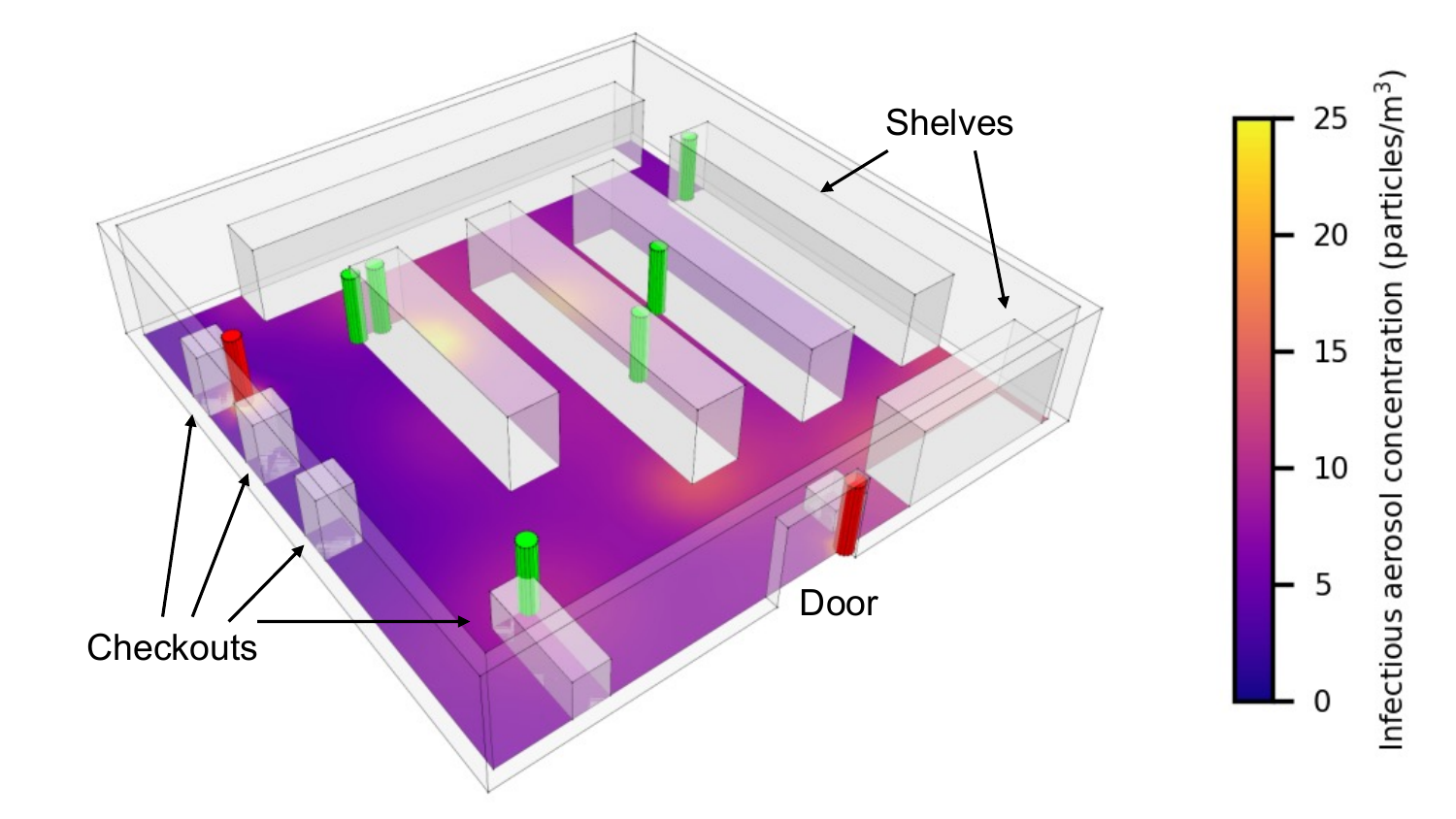}
     \caption{Simulation at 10:00am}
 \end{subfigure}
 \begin{subfigure}[b]{.49\textwidth}
     \centering
     \includegraphics[width=\textwidth]{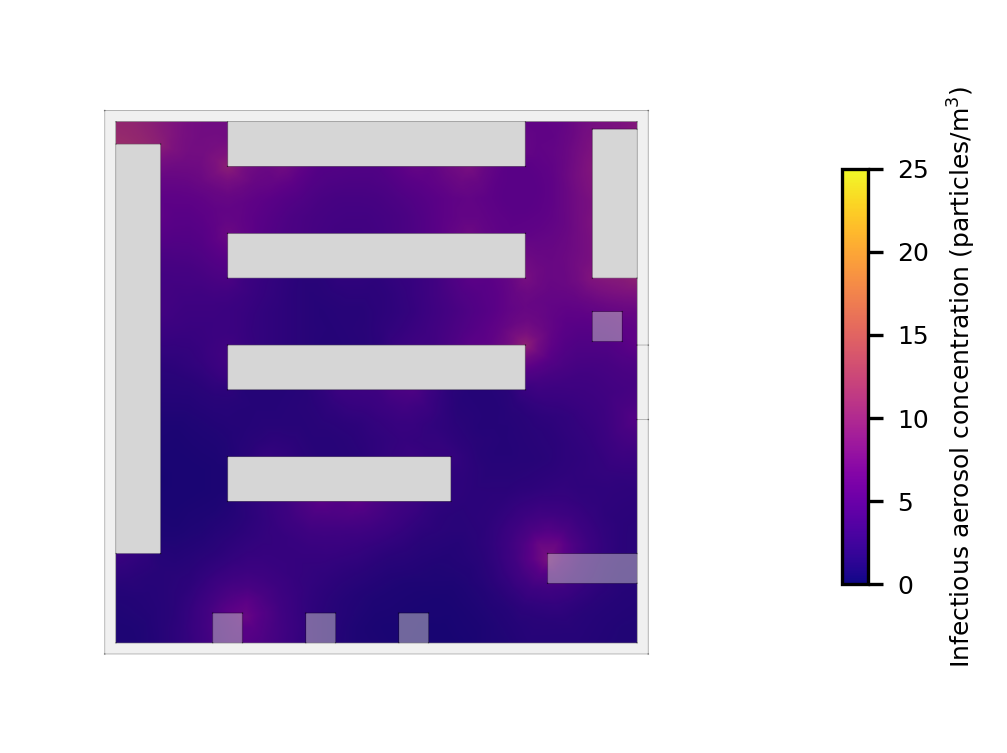}
     \caption{Between 8:00am and 9:00am}
 \end{subfigure}
 \begin{subfigure}[b]{.49\textwidth}
     \centering
     \includegraphics[width=\textwidth]{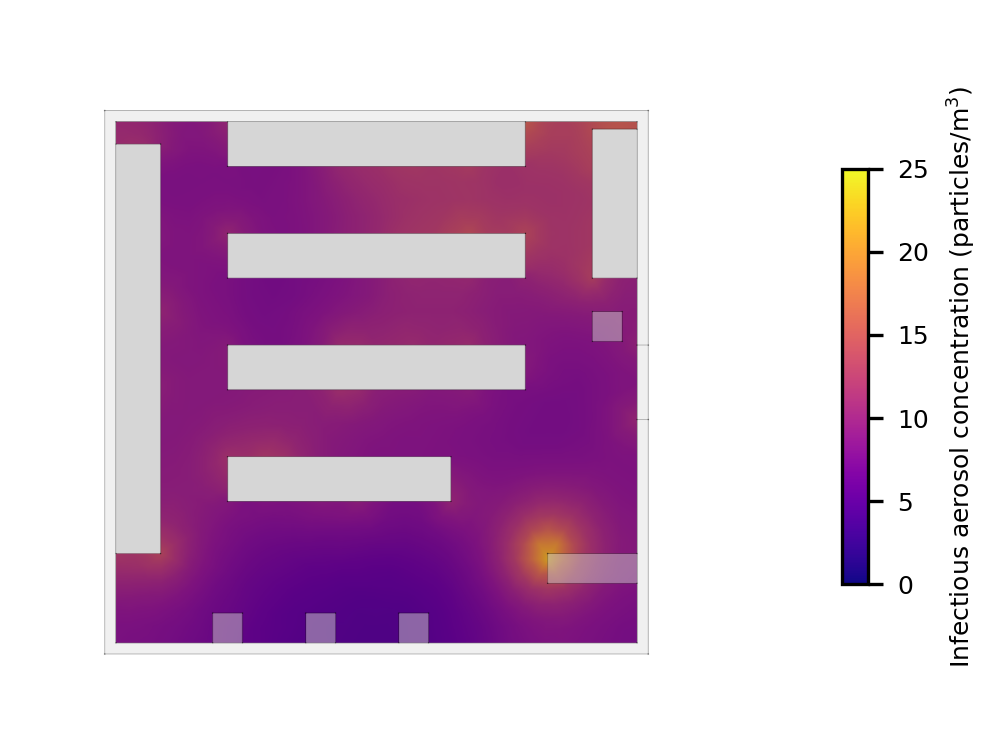}
     \caption{Between 11:00am and 12:00pm}
 \end{subfigure}
  \caption{Supermarket scenario: (a) the concentration of infectious aerosols at 10:00am. Individuals are represented by cylinders (infectious by red colour and susceptible by green colour). The average concentration of infectious aerosols during the (b) first hour (8:00am--9:00am) and (c) last hour of opening (11:00am--12:00pm).}
  \label{fig:supermarket}
\end{figure}

Here, we study a small supermarket and use the model to visualise infection hotspots and again compare several NPIs. The supermarket has dimensions of $14\ \mathrm{m}\ (l)\times14\ \mathrm{m}\ (w)\times3\ \mathrm{m}\ (h)$, as shown in Figure \ref{fig:supermarket}a. We assume that the ventilation is very poor ($\mathrm{ACH}=0.12\ \mathrm{h^{-1}}$, i.e.~$\lambda=3.3{\times}10^{-5}\ \mathrm{s^{-1}}$ and $D=1.9\times10^{-3}\ \mathrm{m^2/s}$).

We generate random schedules for the individuals. Each individual enters the supermarket from the door, picks up a basket, chooses items from the shelves, pays at a checkout and exits through the same door. An individual's schedule is primarily determined by three parameters: frequency of customers entering the supermarket, $f$, the number of items each customer will buy, $N$, and the time it takes to choose each item, $T$. We set $f=1\ \mathrm{min}^{-1}$, $N=6$, and $T$ uniformly distributed from 40 to 80 seconds, that is the average $T$ is 60 seconds and standard deviation is equal to 11.55 seconds. The locations of the six items are uniformly randomly assigned on the boundary of shelves. The middle two checkouts are closed for the original scenario (these two will be opened for a NPI). The customer pays at one of the other two checkouts randomly, when they are not being used. The checkout time is also $T=60{\pm}11.55\ \mathrm{s}$ (defined by a continuous uniform distribution between 40 and 80 s). Each customer, thus, spends an average of 7.5 minutes in the supermarket.

From 8:00am to 12:00pm, 240 individuals, thus, enter the supermarket. As in the care home scenario, we assume that 1/6 of the individuals are infectious (40 individuals are infectious and 200 individuals are susceptible), and that all infectious individuals are not superspreaders ($\mu=6.5\%$). Figure \ref{fig:supermarket}a shows the concentration of infectious aerosols at 10:00am (two hours since the start time). There are two infectious individuals and eight susceptible individuals, represented by red and green colours, respectively. Figures \ref{fig:supermarket}b and \ref{fig:supermarket}c present the average concentration of infectious aerosols in the supermarket during the (b) first and (c) last hour of opening, where it accumulates over time and hotspots are identified in busy aisles and checkout regions.

Next, we compare five interventions: (\RNum{1}) improving ventilation from very poor ($\mathrm{ACH}=0.12\ \mathrm{h^{-1}}$) to good ($\mathrm{ACH}=3\ \mathrm{h^{-1}}$), (\RNum{2}) all customers wear a surgical mask ($\eta=60\%$), (\RNum{3}) halving the number of people entering, (\RNum{4}) reducing the duration of shopping for each customer by 50\% ($T=20$--$40\ \mathrm{s}$), (\RNum{5}) opening the two middle checkouts to avoid crowding. Interventions \RNum{3}, \RNum{4} and \RNum{5} are based on the UK government's guidance for consumers on COVID-19 and food \cite{supermarket_guidance}. We simulate interventions \RNum{1} and \RNum{2} using the original schedule, and we generate new schedules to implement \RNum{3}, \RNum{4} and \RNum{5}.

\begin{figure}
  \centering
  \includegraphics[width=.7\textwidth]{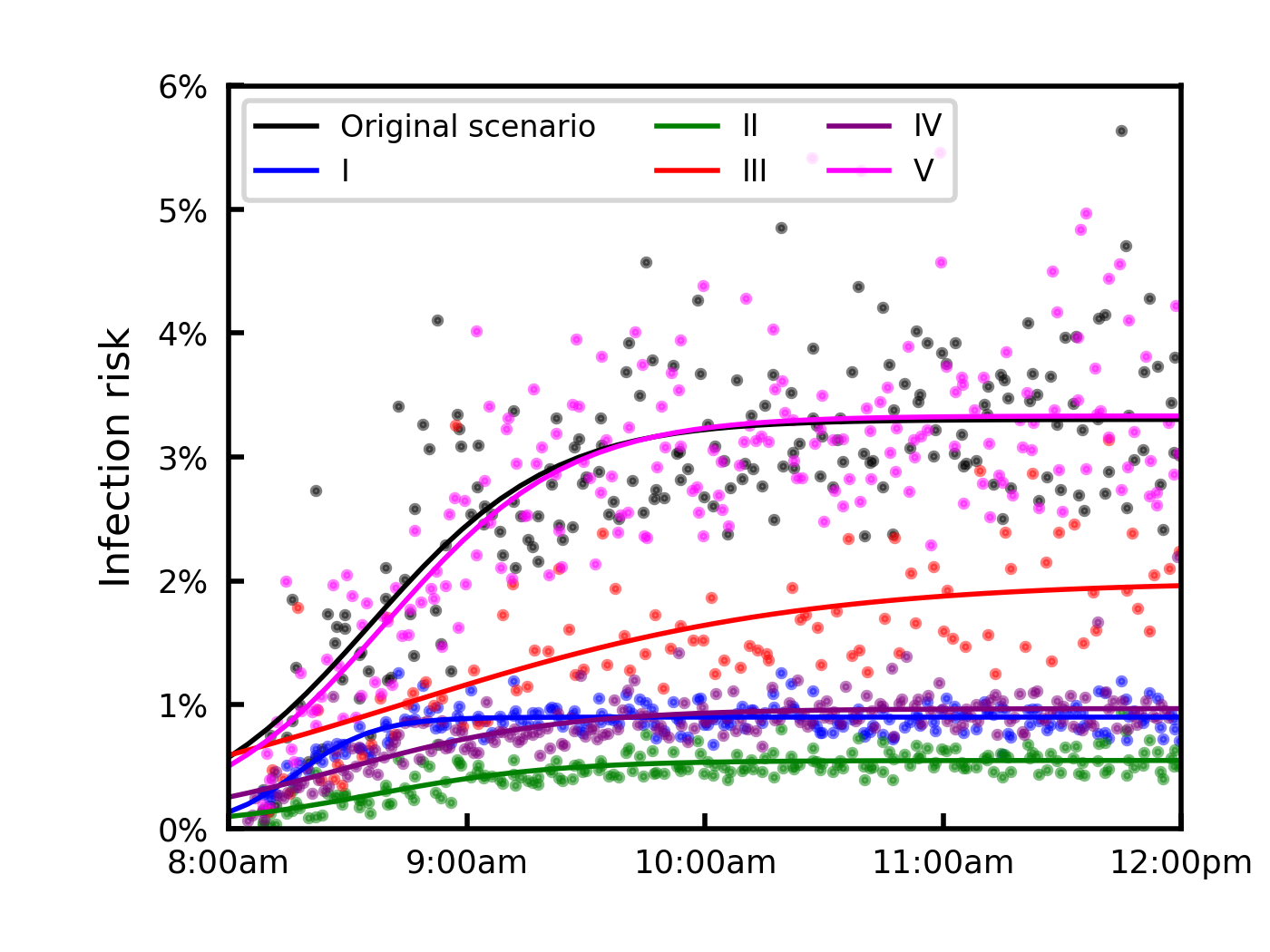}
  \caption{The infection risk of susceptible individuals in the original supermarket scenario, and when implementing different NPIs: (\RNum{1}) improving ventilation from very poor ($\mathrm{ACH}=0.12\ \mathrm{h^{-1}}$) to good ($\mathrm{ACH}=3\ \mathrm{h^{-1}}$), (\RNum{2}) all wearing a surgical mask ($\eta=60\%$), (\RNum{3}) halving the number of people entering, (\RNum{4}) reducing the duration of shopping for each customer by 50\% ($T=30 \mathrm{s}$), (\RNum{5}) opening another two checkouts. The dots represent infection risk of each susceptible individual leaving the supermarket. For each intervention, a logistic curve is fitted to the infection risk data.}
  \label{fig:schedule_comparison_supermarket}
\end{figure}

Figure \ref{fig:schedule_comparison_supermarket} shows the infection risk of susceptible individuals under the original scenario, and when implementing the interventions \RNum{1}--\RNum{5}. Since the supermarket opens, the infection risk increases for approximately 3 hours, due to the accumulation of infectious aerosols, as shown in Figures \ref{fig:supermarket}b and \ref{fig:supermarket}c, then reaches a steady state value during the last hour. For each intervention, the simulation results are fitted to a logistic curve for comparison. Here, we choose the logistic curve, because it exhibits a quasi-linear phase first, followed by a saturation phase, which mimics the observed trend in the infection risk evolution. 

If we compare the infection risk at 12:00pm, the most effective intervention is \RNum{2} (wearing a surgical mask), which reduces the infection risk by 2.8\%. The second most effective intervention is \RNum{1} (improving ventilation to good as $\mathrm{ACH}=3\ \mathrm{h^{-1}}$), which reduces the infection risk by 2.4\%. The third most effective intervention is \RNum{4} (reducing shopping time by 50\%), which reduces the infection risk by 2.3\%, followed by intervention \RNum{3} (halving the number of people entering), reducing the infection risk by 1.3\%. The opening of two more checkouts, however, shows insignificant impact on the infection risk in the scenario considered here.

\subsection{Web app}
\label{sec:web_app}

In this section, we have used VIRIS to study infection spread in three different indoor settings, and demonstrated its ability to assess diverse NPIs. We have also developed a user-friendly web app (\url{https://viris.app}). The app will enable policymakers and space managers to assess NPIs and issue evidence-based guidance, architects to assess architectural designs, and the general public to get a better understanding of NPIs and their relative role in reducing the infection risk.

\begin{figure}
  \centering
  \begin{subfigure}[b]{.7\textwidth}
  \centering
  \includegraphics[width=\textwidth]{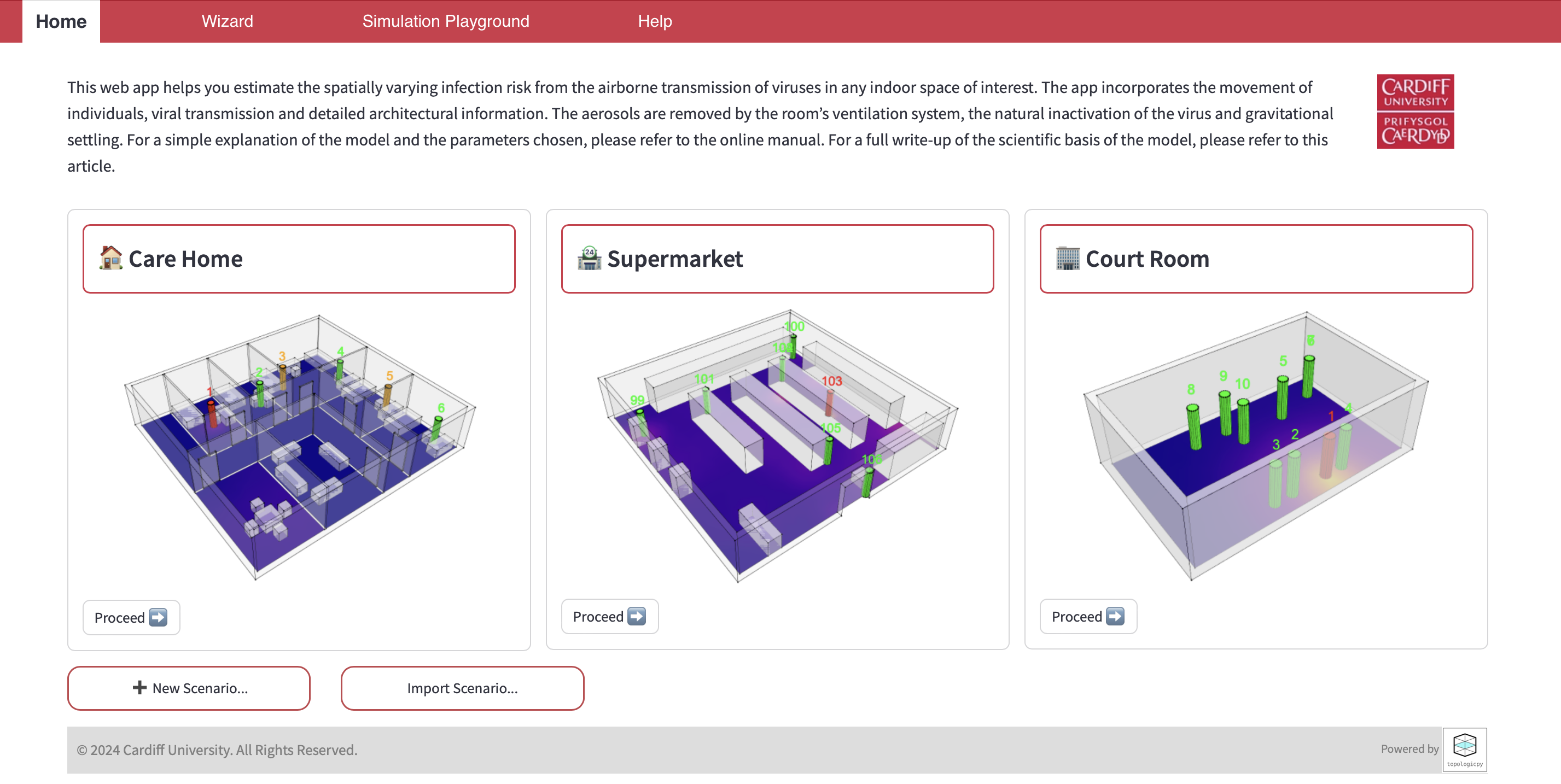}
  \caption{Landing Page}
\end{subfigure}
 \begin{subfigure}[b]{.7\textwidth}
  \centering
  \includegraphics[width=\textwidth]{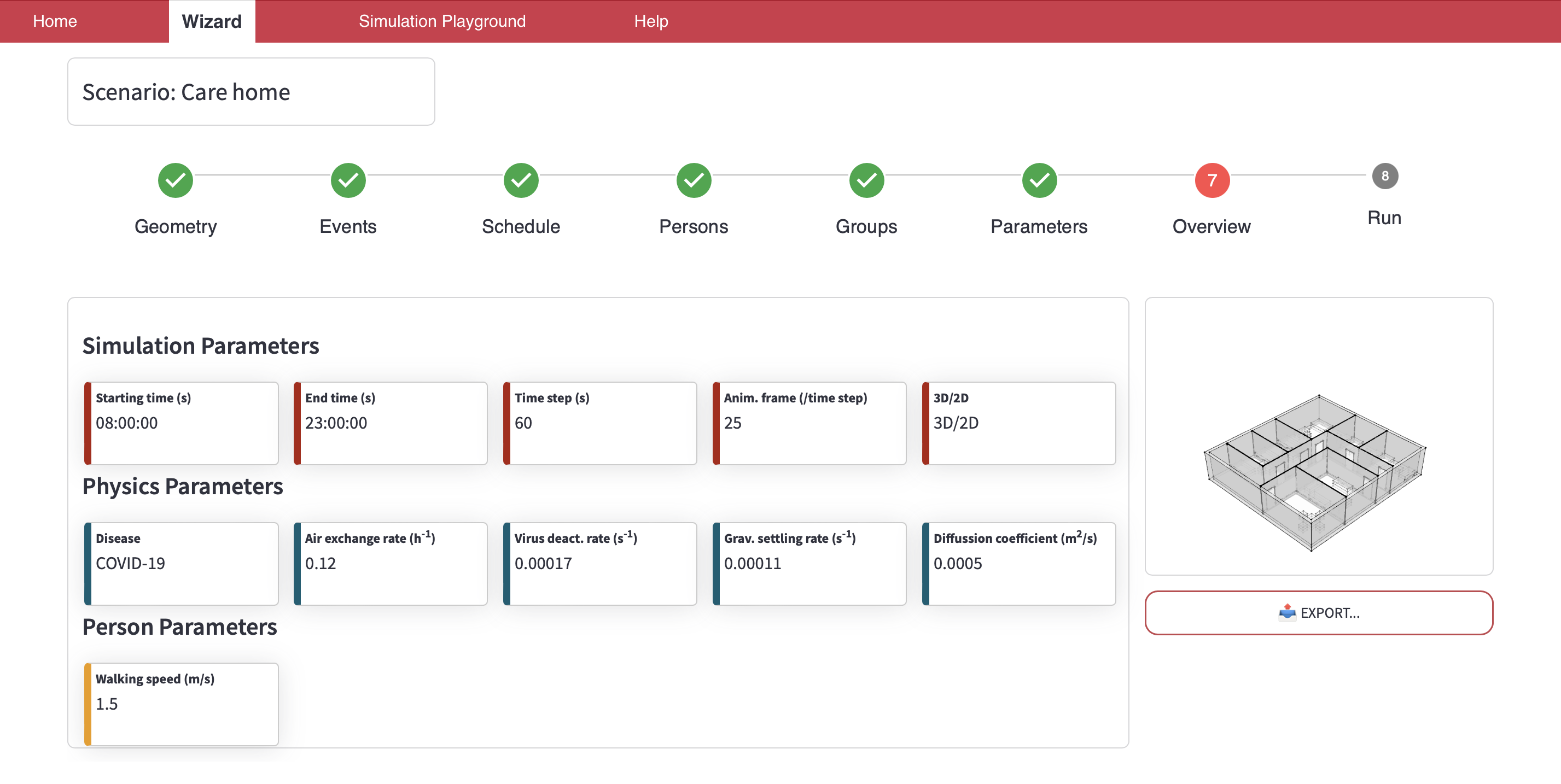}
  \caption{Wizard}
\end{subfigure}
 \begin{subfigure}[b]{.7\textwidth}
  \centering
  \includegraphics[width=\textwidth]{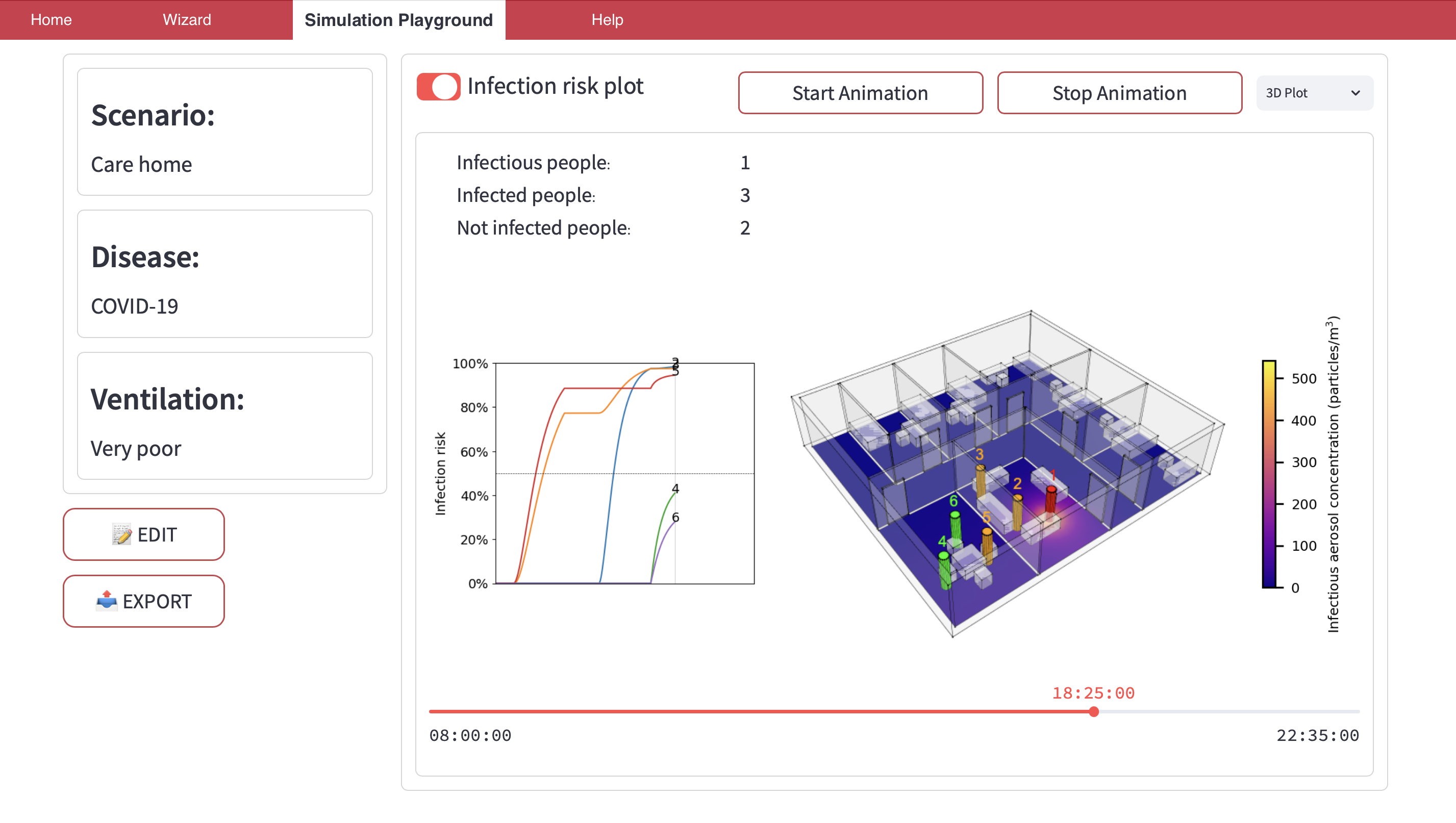}
  \caption{Simulation Playground}
\end{subfigure}
  \caption{Screenshots of the web app: a) Landing Page, b) Wizard for setting up the simulation, and c) Simulation Playground for visualising the results.}
  \label{fig:web_app}
\end{figure}

Figure \ref{fig:web_app} shows three screenshots of the VIRIS app. Figure \ref{fig:web_app}a shows the Landing Page. Figure \ref{fig:web_app}b presents the Overview panel of the Wizard to set up a simulation. The wizard allows the user to set up the simulation step-by-step: the users can import their own architectural geometry or select from preset scenarios, specify people movements, and set parameter values. The simulation is executed at the last step pressing ``Run'', and the simulation results are shown on the Simulation Playground, as shown in Figure \ref{fig:web_app}c. At the Simulation Playground, the users can export their simulation results (including raw data) to their local machine. The user can visualise the simulation as an animation (3D or 2D), see a summary table and also a plot of the individuals' infection risk over time. The users can also use the scroll bar on the bottom to visualise a specific time step of the simulation.

The app has been discussed with the Welsh Government and we are looking into using it to study various indoor settings according to the government's priorities.

\section{Discussion}
\label{sec:discussion}

In this paper, we present VIRIS, a new epidemic simulator that combines airborne viral transmission, people movement, and detailed architectural design\cite{topologicpy,Wassim2024}. The modelling framework couples airborne transmission modelling based on the Wells-Riley ansatz \cite{Wells1955,Riley1978,Lau2022} with agent-based simulations of individuals' movement, and the architectural software topologicpy \cite{topologicpy,Wassim2024}. VIRIS enables fast simulations employing a finite element method to solve the reaction-diffusion equation (\ref{eq:governing_equation}) for the viral concentration; a simulation of tens of individuals moving over a whole day can be computed within 1 minute on a standard laptop. The model's versatility and applicability has been demonstrated in three indoor settings: courtroom, care home and supermarket.

The model aims to enable policymakers and space managers to assess and issue guidance to mitigate the transmission of airborne diseases in indoor spaces. In addition, it could help architects and engineers to assess designs that mitigate infection risk \cite{Wassim2024}. In this work, we examine and rank different NPIs, many of those included in UK Government guidance, for care homes \cite{care_home_guidance} (Figure \ref{fig:schedule_comparison_care_home}) and for a small supermarket \cite{supermarket_guidance} (Figure \ref{fig:schedule_comparison_supermarket}).

Three NPIs have been studied for both the care home and the supermarket: improving ventilation, wearing a surgical mask, and reducing the duration of person-person interactions. Based on our simulations, wearing a surgical mask (60\% efficiency) has been found to be the most effective NPI in both cases. Improving ventilation from very poor ($\mathrm{ACH}=0.12\ \mathrm{h^{-1}}$) to good ($\mathrm{ACH}=3\ \mathrm{h^{-1}}$) is the second most effective intervention in both scenarios. However, improving ventilation from very poor to poor ($\mathrm{ACH}=0.72\ \mathrm{h^{-1}}$) appears to be less effective in both scenarios. 

In addition, we highlight the importance of designing and studying scenario-customised NPIs. For example, we consider socialising within small groups in the care home, and show that it is the third most effective intervention following imposing no common room time and everyone wearing a surgical mask. In the supermarket, we instead study halving the number of customers and opening two more checkouts. The former intervention greatly reduces the infection risk, while the latter one is found to be ineffective. These results underscore the need to assess interventions specific to each setting, a task our model can easily fulfil.

We emphasise that all results presented here are specific to the parameters, indoor layout, and schedules used in this paper. These are used to showcase the ease of customising our model to evaluate diverse NPIs, while the quantitative results and conclusions need further validation and verification before being used to inform real-world decisions. Change of parameters can be very easily done in VIRIS and this is further facilitated by the VIRIS app.

The care home and the supermarket are studied based on typical but fictitious designs \cite{Khaliq2024}; model calibration using real geometries is needed before using VIRIS to inform decision-making in real life. Model calibration for various real settings is in development through our collaboration with policymakers, space managers and architects. Moreover, extensions in the modelling framework can be undertaken; for instance, to consider multi-level buildings and to solve equation (\ref{eq:governing_equation}) in three dimensions. These extensions can be well supported by topologicpy \cite{topologicpy,Wassim2024}, but require substantial changes in the simulator and the web app. Future work could also incorporate more detailed fluid dynamics models \cite{Mingotti2020,Cui2021,Zhen2022,Jia2022,Lau2022,Pretty2023,Lim2024}, but preserving computational efficiency would then be very challenging.

In conclusion, we have presented VIRIS, a modelling framework, simulator and app for airborne viral transmission, which combines detailed people movement and architectural design. We have demonstrated the applicability and versatility of VIRIS by studying several NPIs in a care home and a supermarket.

\bibliography{bibliography}

\section*{Acknowledgements}

This work is funded by an EPSRC Impact Acceleration Account grant, under grant number EP/X525522/1. We would like to acknowledge fruitful discussions with Dr Rob Orford and Dr Brendan Collins, while they held, respectively, the posts of Chief Scientific Advisor (Health) and Head of Health Economics, Advanced Analysis and Policy Modelling in the Welsh Government.

\section*{Author contributions statement}

YX: Conceptualization, Methodology, Software, Validation, Formal Analysis, Data Curation, Writing – Original Draft, Visualization.
WJ: Methodology, Software, Writing – Review \& Editing, Visualization, Supervision, Funding Acquisition.
TEW: Conceptualization, Methodology, Formal Analysis, Writing – Review \& Editing, Supervision, Funding Acquisition.
KK: Conceptualization, Methodology, Formal Analysis, Writing – Review \& Editing, Supervision, Project Administration, Funding Acquisition.

\end{document}